\documentclass[a4paper,12pt]{article}\newif\ifpreprint\preprinttrue
\usepackage{amsmath,amssymb}%
\usepackage{txfonts}
\usepackage[sort&compress,super,comma]{natbib}
\usepackage{graphicx}
\newdimen{\figbase}
\newdimen{\textbase}
\ifpreprint
  \newcommand{\titlefontsize}{\normalsize}
\else
  \newcommand{\titlefontsize}{\large}
\fi

\newcommand{\etal}{{\it et al}}
\newcommand{\Tc}{$T_c$}
\newcommand{\Tconset}{$T_c^{\mbox{\scriptsize onset}}$}
\newcommand{\Tczero}{$T_c^0$}
\newcommand{\tsp}{$t_{\mbox{\scriptsize sp}}$}
\newcommand{\affCrystMaterSci}{\titlefontsize%
  Department of Crystalline Materials Science, %
  Nagoya University,\\
  \titlefontsize%
  Chikusa-ku, Nagoya 464-8603, Japan
}
\newcommand{\affVBL}{\titlefontsize%
  Venture Business Laboratory (VBL), %
  Nagoya University,\\
  \titlefontsize%
  Chikusa-ku, Nagoya 464-8603, Japan
}
\newcommand{\affTRIP}{\titlefontsize%
  JST, Transformative Research-Project on %
  Iron Pnictides (TRIP),\\ 
  \titlefontsize
  Sanbancho 5, Chiyoda-ku, Tokyo 102-0075, Japan
}

\makeatletter
\renewcommand\section{\@startsection {section}{1}{\z@}%
                                   {-3.5ex \@plus -1ex \@minus -.2ex}%
                                   {2.3ex \@plus.2ex}%
                                   {\normalfont\bfseries}}
\makeatother

\begin{document}

\setlength{\baselineskip}{\textbase}

\title{\titlefontsize\textbf{
  In-situ growth of superconducting NdFeAs(O,F) thin films 
  by Molecular Beam Epitaxy
  }
}

\author{\titlefontsize
  T.\ Kawaguchi$^{1,3*}$, H.\ Uemura$^{1,3}$, T.\ Ohno$^{1,3}$, 
  M.\ Tabuchi$^{2,3}$, T.\ Ujihara$^{1,3}$,\\
  \titlefontsize
  K.\ Takenaka$^{1,3}$, Y.\ Takeda$^{1,3}$ and H.\ Ikuta$^{1,3}$\\\\
  $^1$\affCrystMaterSci\\
  $^2$\affVBL\\
  $^3$\affTRIP
}

\date{\titlefontsize\today}


\maketitle

\textbf{
The recently discovered high temperature superconductor F-doped 
LaFeAsO \cite{Kamihara} 
and related compounds 
\cite{GFChen,RenCPL,RenEPL,Kito,Rotter,Tapp,Hsu,Zhu,Sato}
represent a new class of superconductors with 
the highest transition temperature (\Tc) apart from the cuprates.
The studies ongoing worldwide are revealing that 
these Fe-based superconductors are forming a unique class of materials
that are interesting from the viewpoint of applications. 
To exploit the high potential of the Fe-based superconductors
for device applications,
it is indispensable to establish a process 
that enables the growth of high quality thin films.
Efforts of thin film preparation started soon after 
the discovery of Fe-based superconductors 
\cite{Hiramatsu1111,Hiramatsu122,Backen},
but none of the earlier attempts had succeeded in 
an \textit{in-situ} growth of a superconducting film 
of \textit{Ln}FeAs(O,F) (\textit{Ln}=lanthanide),
which exhibits the highest \Tc\ to date among 
the Fe-based superconductors.
Here, we report on the successful growth of 
NdFeAs(O,F) thin films on GaAs substrates,
which showed well-defined superconducting transitions up to 48 K
without the need of an ex-situ heat treatment.
}

Thin film preparation of 
Co-doped $AE$Fe$_2$As$_2$ ($AE$=Sr, Ba) \cite{Hiramatsu122,Katase2009,Iida,Lee} 
and iron-chalcogens \cite{Han,MJWang,Mele,Bellingeri,Imai}
have been reported from many different research groups,
the best films of which already possessing a \Tc\ value 
exceeding 20 K.
Quite recently, the fabrication of K-doped BaFe$_2$As$_2$
with an onset \Tc\ up to 40 K was also reported \cite{LeeK122}. 
In contrast, the growth of thin films of $Ln$FeAsO, 
the so-called 1111 family, 
which possesses \Tc\ values up to 55 K,
has been confronted with more difficulties.
Hiramatsu \etal.\ prepared La-1111 films 
by pulsed-laser deposition (PLD) on oxide crystalline 
substrates \cite{Hiramatsu1111}. 
They have succeeded in obtaining epitaxial films,
but the films did not show a superconducting transition. 
Soon thereafter, 
another group reported the growth of 
F-doped La-1111 that exhibited a superconducting transition,
which was also prepared by PLD on oxide substrates \cite{Backen,Haindl,Kidszun}.
However, it was necessary to treat
these films ex-situ at significantly high temperatures 
to form the 1111 phase.
The necessity of such heat treatment places a considerable burden 
on the processing of superconducting devices,
and the development of a procedure 
that does not require a subsequent high-temperature treatment
is a necessary next step.

While the above mentioned two groups had employed PLD methods,
we have succeeded in growing Nd-1111 thin films 
by molecular beam epitaxy (MBE) in a previous work \cite{Kawaguchi}. 
The films were grown on GaAs substrates,
which also differentiates our work from the other studies. 
GaAs was selected because of the good lattice matching and 
the similarity in the atomic coordination around As with \textit{Ln}-1111. 
We also expect that various growth techniques established 
in studies on semiconductors can be applied 
when GaAs is used as substrates, 
such as strain control by an appropriate alloy buffer, 
and that novel devices may be realized by combining 
a superconductor with a semiconductor. 

The growth condition for obtaining a single-phased film of 
Nd-1111 with a very high reproducibility was identified from 
a detailed study in our previous work \cite{Kawaguchi}. 
However, the film did not show a superconducting transition
and the resistivity increased at low temperature. 
Because that film was only 15 nm thick,
we speculated that it might have some structural imperfections
and studied the effect of increasing the film thickness.
As a result, we succeeded in obtaining 
as-grown, superconducting thin films of F-doped Nd-1111. 
The highest onset temperature \Tconset\ of these films 
is 48 K with a zero resistance temperature \Tczero\ of 42 K. 
For the best of our knowledge, 
these are the first 1111-phase thin films
that showed superconducting transitions
without the need of an ex-situ heat treatment,
and the \Tc\ values are the highest among all 
thin films of the Fe-based superconductors
reported so-far.

\begin{figure}
  \centerline{\includegraphics[width=\figbase]{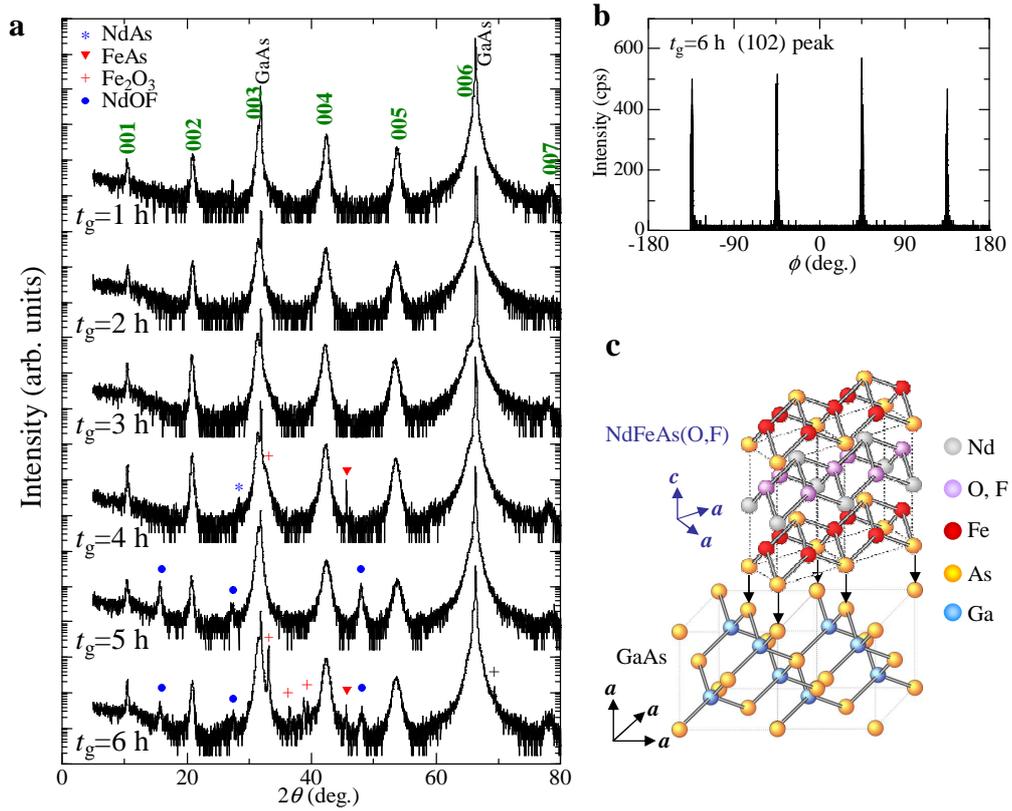}}
  \caption{\label{fig:XRD}
    \textbf{
      X-ray characterization of the Nd-1111 films.
    }
    \textbf{a}, Out-of-plane $\theta$-2$\theta$ XRD profiles of films grown 
    with various growth times $t_g$.
    The growth parameters were identical for all films except 
    the growth time.
    These profiles indicate that the Nd-1111 phase 
    was grown with the $c$-axis perpendicular 
    to the substrate in all films.
    \textbf{b}, Azimuthal phi-scan profile of the off-axis 
    (102) reflection of the $t_g$=6 h film.
    A clear fourfold symmetry can be confirmed. 
    \textbf{c}, A schematic drawing of the epitaxial relationship
    between the Nd-1111 phase and the GaAs substrate,
    which is consistent with the results of XRD analyses.
  }
\end{figure}

Figure \ref{fig:XRD}a shows out-of-plane $\theta$-2$\theta$ XRD patterns of 
the Nd-1111 films grown with various growth time $t_g$. 
The optimal growth condition of our 
previous study \cite{Kawaguchi} was employed except 
for the substrate temperature, 
which was slightly lowered to 650$^\circ$C in an attempt 
to reduce the impurities that were observed in films 
grown with long $t_g$ (see below). 
Except for the growth time, 
all other parameters were identical for the films shown 
in Fig.\ \ref{fig:XRD}a. 
As can be seen, the films grown for $t_g\leq3$ h were single-phased
and all XRD peaks could be indexed 
as (00$l$) reflections from Nd-1111
except those arising from the substrate.
With increasing the growth time further, 
however, the growth became somewhat unstable 
and some impurity peaks were observed in the XRD patterns.
In particular, the formation of NdOF phase is obvious
in the films grown for $t_g$=5 and 6 h.
A fourfold symmetry was confirmed 
for the Nd-1111 phase in all films 
by measuring the azimuthal phi-scan of the (102) peak, 
as shown for example for the $t_g$=6 h film in Fig.\ \ref{fig:XRD}b. 
These results indicate that the Nd-1111 phase 
was grown epitaxially on the substrate
with the relation schematically shown in Fig.\ \ref{fig:XRD}c.

\begin{figure}
  \centerline{\includegraphics[width=0.7\figbase]{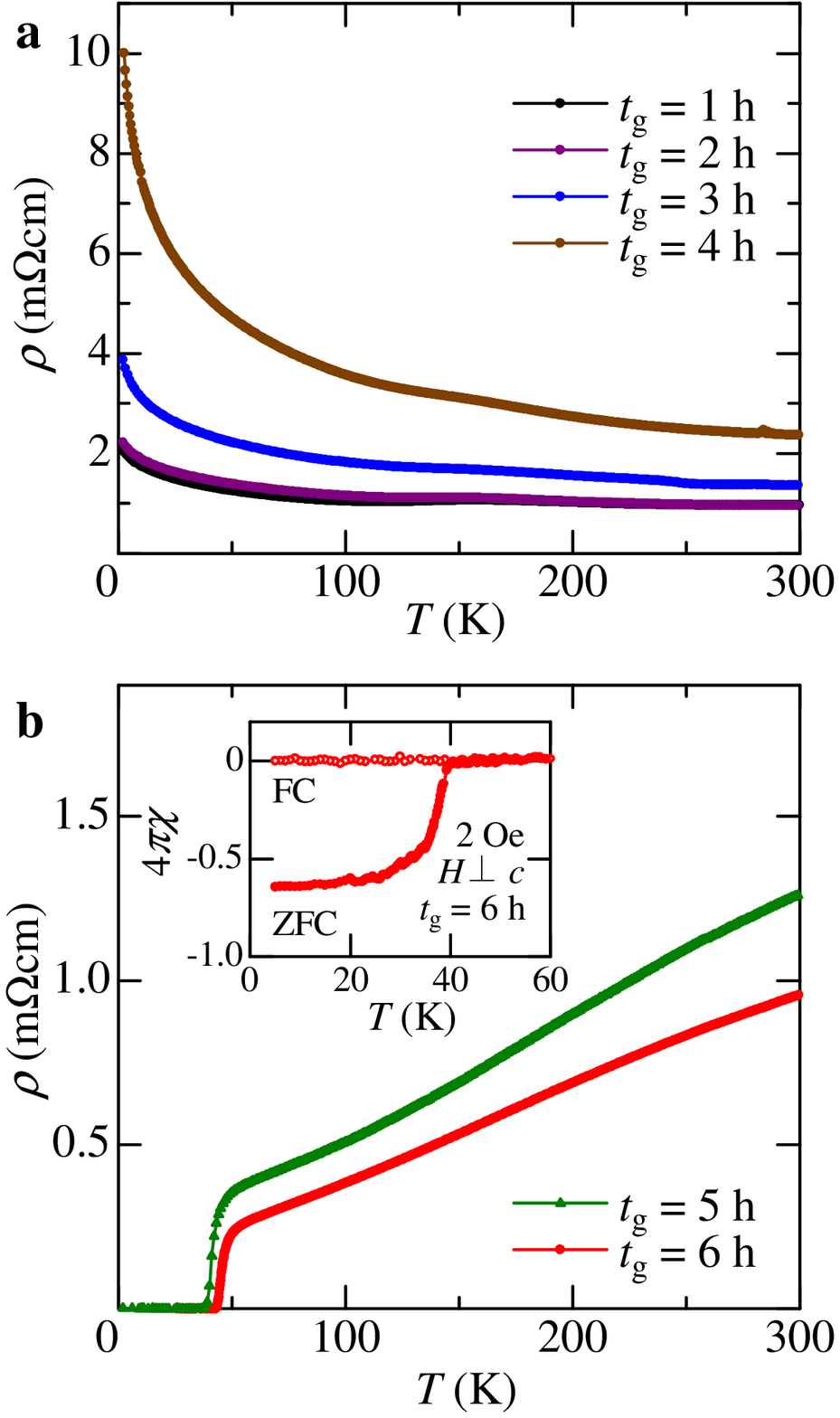}}
  \caption{\label{fig:rchi}
    \textbf{
      Temperature dependence of resistivity of 
      the Nd-1111 films.
    }
    \textbf{a}, $\rho(T)$ curves of the films grown for $t_g\leq4$ h.
    \textbf{b}, $\rho(T)$ curves of the films grown for 
    $t_g$=5 and 6 h.
    The films grown for $t_g\geq5$ h exhibited 
    clear superconducting transitions
    with \Tconset=45 K and \Tczero=38 K for the $t_g$=5 h film and  
    \Tconset=48 K and \Tczero=42 K for the $t_g$=6 h film. 
    The inset shows the temperature dependence of susceptibility of 
    the $t_g$=6 h film.
  }
\end{figure}

Figure \ref{fig:rchi} shows the temperature dependence 
of resistivity ($\rho(T)$) of the Nd-1111 films.
The film thickness that was necessary for converting
resistance data to resistivity
was calculated on the basis of 
the previously determined growth 
rate 15 nm/h (ref.\ 25), 
with the assumption that it did not depend on the growth time. 
As shown in Fig.\ \ref{fig:rchi}a,
the resistivity of the films grown for $t_g\leq4$ h 
increased at low temperature similarly to 
the film we reported in our previous study \cite{Kawaguchi}
and did not show a superconducting transition. 
In a stark contrast, however,
the $t_g$=5 and 6 h films shown in Fig.\ \ref{fig:rchi}b
exhibited a superconducting transition. 
\Tconset\ and \Tczero\ of the $t_g$=6 h film were 
slightly higher than the $t_g$=5 h film,
and were 48 K and 42 K, respectively. 
The susceptibility of the $t_g$=6 h film, 
which is shown in the inset to Fig.\ \ref{fig:rchi}b, 
confirms also a superconducting transition at about 40 K. 

\begin{figure}
  \centerline{\includegraphics[width=0.7\figbase]{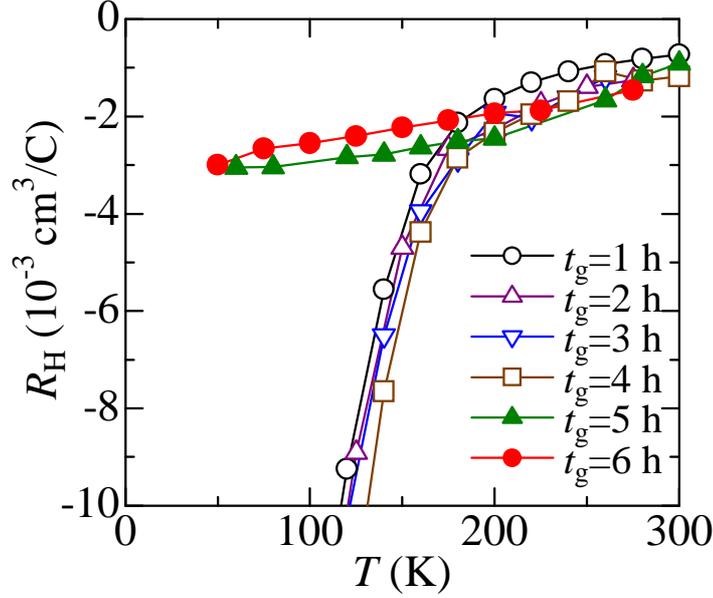}}
  \caption{\label{fig:Hall}
    \textbf{
      Temperature dependence of Hall coefficient of 
      the Nd-1111 films.
    }
    A rather sudden change in the behavior of Hall coefficient 
    was observed between $t_g$=4 and 5 h,
    which corresponds quite well to the difference between 
    the reported data of non-doped and F-doped bulk samples.
 }
\end{figure}

The difference between the $t_g\leq4$ h and $t_g\geq5$ h films 
can be attributed to F-contents. 
Figure \ref{fig:Hall} shows the temperature dependence of 
Hall coefficient of the Nd-1111 films. 
All films exhibited a negative Hall coefficient 
for the temperature range investigated. 
The magnitude of Hall coefficient of the $t_g\leq4$ h films, 
which did not show a superconducting transition,
increased steeply below about 200 K. 
On the other hand,  the Hall coefficient of the $t_g$=5 and 6 h films
had a much weaker temperature dependence.
This change in the behavior of Hall coefficient
corresponds quite well to the difference between non-doped and F-doped 
single crystals of Nd-1111 \cite{Cheng},
suggesting that the films grown for $t_g$=5 and 6 h 
were successfully doped with fluorine. 
Interestingly, the change in the behavior of Hall coefficient 
took place rather suddenly
between $t_g$=4 and 5 h.
The results of electron probe micro-analysis (EPMA) is consistent 
with this observation
because no fluorine was found in 
the $t_g\leq4$ h films while the $t_g$=5 and 6 h 
films contained a certain amount of fluorine.

\begin{figure}
  \centerline{\includegraphics[width=0.8\figbase]{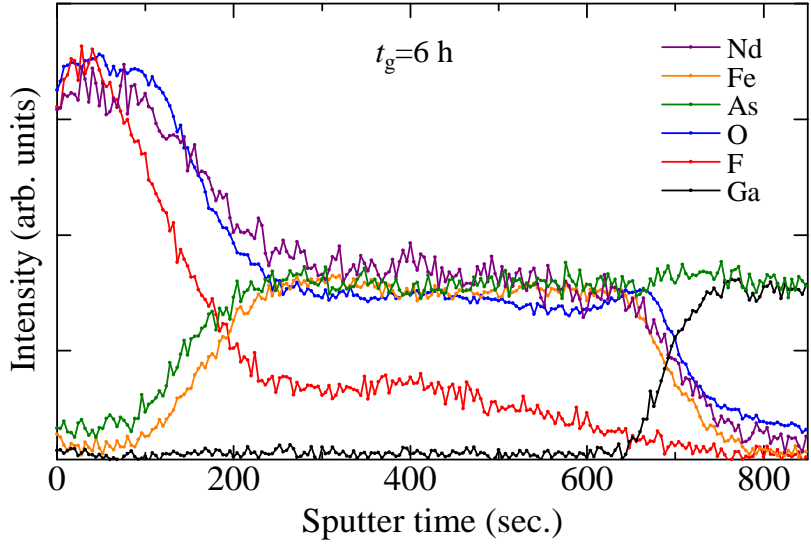}}
  \caption{\label{fig:AES}
    \textbf{
      Auger depth-profile analysis of the $t_g$=6 h Nd-1111 film.
    }
    The intensity of the Auger signal is plotted as a function of 
    sputtering time.
    The purple, orange, green, blue, red, and black curves
    correspond to Nd, Fe, As, O, F, and Ga, respectively.
  }
\end{figure}

Next, we would like to discuss why fluorine was 
observed only in the films that were grown with long $t_g$.
Figure \ref{fig:AES} shows the Auger depth profile 
of the $t_g$=6 h film, 
which is a plot of the intensity of the Auger signals 
as a function of the sputtering time \tsp.
The interface between the film and the substrate 
is evidenced by a rise in the Ga signal,
and it can be seen that the contents of 
Nd, Fe, As, and O are nearly constant in the region above the interface 
for a certain range of thickness.
The presence of fluorine was also confirmed,
which is consistent with the conclusion of Hall coefficient
and EPMA  measurements.
Very interestingly, however, 
we observed steep increases in Nd, O, and F contents
near the surface (\tsp\ $\lesssim200$ s) accompanied 
with depletion of Fe and As contents.
This implies that the NdOF phase, 
that was observed in the XRD analyses of 
the $t_g$=5 and 6 h films,
was formed at the end of the film growth. 
Indeed, the reflection high-energy electron diffraction (RHEED)
pattern that was monitored during the growth 
suggested that the Nd-1111 phase was grown until $t_g\sim4$ h, 
while a different phase, presumably NdOF, 
was dominant for $t_g\gtrsim4$ h. 

The change in the dominantly growing phase from 
Nd-1111 to NdOF at $t_g\simeq4$ h
was entirely unexpected for us
because all processing parameters were kept to the same.
However, we point out that the films were grown in an atmosphere 
that was probably quite excessive in fluorine. 
This is because NdF$_3$ was used as the source of Nd and F,
which means that the supplied amount of 
F was three times larger than Nd.
Therefore, what was unusual might be not the formation of NdOF 
but rather the Nd-1111 phase during the first stage
of the film growth.
It has been reported that fluorine can react with GaAs 
forming GaF$_3$ when GaAs is exposed to a F-containing vapor
and the formed GaF$_3$ sublimates above about 550 K 
(about 280$^\circ$C) \cite{Freedman,Simpson}.
We think that the same reaction took place
at the early stage of the film growth,
and the GaF$_3$ phase had immediately sublimated
because the substrate temperature was 650$^\circ$C.
This consequently had an effect of regulating 
the amount of fluorine, 
and the Nd-1111 phase had grown.
With the increase in the film thickness, however, 
reaching the GaAs surface became increasingly difficult for fluorine,
and some of the fluorine remained unconsumed.
When the amount of fluorine exceeded then a certain level, 
the growth of NdOF was thermodynamically more favorable,
causing the change in the dominantly growing phase.
SEM observations indicated that there exist some
holes beneath the Nd-1111 films,
which were absent when elemental Nd was used instead
of NdF$_3$ as one of the sources.
We think that these holes were formed due to the reaction of fluorine
with the substrate,
and support our growth model.

The present model can explain why 
the dominantly growing phase
changed with the growth time. 
However, one would expect then that 
a F-doped Nd-1111 film with no NdOF should be obtained 
when the growth is stopped at $t_g\sim4$ h.
Nevertheless, the results of Hall coefficient and EPMA measurements
indicate that this is not the case.
A possible explanation for the lack of fluorine 
in the $t_g\leq4$ h films
is that fluorine may diffuse easily
through Nd-1111 at high temperature 
and had escaped from the film after 
the supply of NdF$_3$ was stopped to terminate the growth.
The NdOF phase that was grown in the $t_g\geq5$ h films
may have worked as a cap layer
and had prevented the loss of fluorine.
In this respect, it is interesting that 
Kidszun \etal.\ had mentioned that 
the formation of LaOF at the surface 
is a typical feature in their F-doped La-1111 films \cite{Kidszun}. 

In summary, we have successfully grown superconducting 
films of Nd-1111.
The as-grown films exhibited superconducting transitions 
when the growth time was sufficiently long
with \Tconset\ and \Tczero\ up to 48 K and 42 K, respectively.
While the films grown for $t_g\leq3$ h were single-phased, 
we found a NdOF layer near the surface in the films 
that exhibited superconducting transitions. 
We think that this is because 
our films were grown in an excess supply of fluorine,
but the presence of the cap layer may have played
an important role to realize an as-grown superconducting film
in the present work.
This conclusion points some directions to which 
future studies should move to realize
a film that is both single-phased \textit{and} superconducting.
First, it is preferable to avoid the reaction 
of fluorine with the substrate
because it changes the amount of fluorine 
that contributes to the film formation 
and complicates the process.
This may be achieved by using an appropriate buffer layer 
that protects GaAs.
An independent control of Nd and F is also necessary
because the excess supply of F 
seems to favor the formation of NdOF.
Finally, the loss of fluorine from the film
has to be prevented, 
which may be achieved by lowering the growth temperature.
Studies in these directions are underway. 

\vspace{\baselineskip}
\noindent{\bf Methods}

\noindent
The thin films were grown by a MBE method, 
the detail of which is reported in 
the previous paper \cite{Kawaguchi}. 
Briefly, GaAs(001) was used as substrates, 
on which an about 300-nm-thick GaAs buffer layer was grown 
at 610$^\circ$C after the oxide layer on the substrate 
was removed by sublimation. 
Nd-1111 was grown by supplying all elements 
from solid sources charged in Knudsen cells; 
Fe, As, NdF$_3$, and Fe$_2$O$_3$. 
The substrate temperature was 650$^\circ$C and 
the vapor pressures of Fe, As, NdF$_3$, and oxygen were 
$1.9\times10^{-6}$, $1.5\times10^{-3}$, 
$2.7\times10^{-6}$, and $2\times10^{-5}$ Pa, respectively.
The composition of the films was checked
by electron probe micro-analysis (EPMA) using 
Nd-1111 powders as a reference. 
Depth-profile analysis was performed using
an Auger electron spectroscopy (AES)
combined with Ar ion sputtering (JEOL JAMP-7800).
Scanning electron microscope (SEM) observation was performed
using HITACHI S-4300.
The growth rate of the films was estimated to be 
15 nm/h in our previous study \cite{Kawaguchi}. 
Resistivity was measured by a four-probe method, 
and susceptibility was measured using 
a SQUID magnetometer (Quantum Design MPMS-7). 
Hall coefficient was measured under 
a magnetic field of 9 T.

\vspace{2\baselineskip}
\noindent{\bf Acknowledgement}

\noindent
This work was supported by 
Transformative Research Project on Iron Pnictides (TRIP), 
Japan Science and Technology Agency (JST).

\end{document}